\pdfoutput=1
\documentclass[conference]{IEEEtran}
\IEEEoverridecommandlockouts

\usepackage[normalem]{ulem}
\usepackage{cite}
\usepackage{amsmath,amssymb,amsfonts}
\usepackage{graphicx}
\usepackage{url}   
\usepackage{subcaption} 
\usepackage{graphicx}
\usepackage{subcaption}
\usepackage{caption}
\usepackage{float}
\usepackage{textcomp}
\usepackage{xcolor}
\usepackage{CJKutf8}  
\usepackage{enumerate}
\usepackage{enumitem} 
\usepackage[ruled,vlined,linesnumbered]{algorithm2e}
\usepackage{setspace}
\usepackage{bm}
\usepackage{accents}
\usepackage{booktabs}   
\usepackage{multirow}   
\usepackage{siunitx}    
\def\BibTeX{{\rm B\kern-.05em{\sc i\kern-.025em b}\kern-.08em
    T\kern-.1667em\lower.7ex\hbox{E}\kern-.125emX}}
\newcommand{\hvec}[1]{\accentset{\rightharpoonup}{#1}}
\newcommand{\TT}{\text{TT}}

\newcommand{\CBS}{\text{CBS}}
\newcommand{\idSl}{\text{idSl}}
\newcommand{\sdSl}{\text{sdSl}}

\newcommand{\myconcept}[1]{\textsc{#1}}

\newtheorem{definition}{Definition}

\begin{document}
\title{DeepNC: A Fast GNN-based Pre-Verification Surrogate for TSN Configuration}

\author{
\IEEEauthorblockN{Jiayi Zhu\textsuperscript{a}$^{\dagger}$, Jing Lin\textsuperscript{a}$^{\dagger}$, Zelong Tian\textsuperscript{a}, Feng He\textsuperscript{a}, Luxi Zhao\textsuperscript{a}\textsuperscript{*}}
\IEEEauthorblockA{\textsuperscript{a}School of Electronics and Information Engineering, Beihang University, China}
\thanks{$^{\dagger}$ Co-first authors with equal contribution. $^{*}$Corresponding author.}
\thanks{Emails: zhujiayi@buaa.edu.cn; linjing1023@buaa.edu.cn; zelongtian@buaa\protect\\.edu.cn; fenghe@buaa.edu.cn; zhaoluxi@buaa.edu.cn.}
}
\maketitle

\begin{abstract}
Time-Sensitive Networking (TSN) is critical to deterministic communication in safety-critical domains, with formal verification such as Network Calculus (NC) serving as the cornerstone for schedulability guarantees. However, during automated configuration-space exploration, repeated schedulability analysis consumes over 90\% of the total configuration time, becoming the primary bottleneck for large-scale TSN configurations. To address this challenge, we propose DeepNC, a novel pre-verification surrogate module that pioneers the structural fusion of NC principles into a Graph Neural Network (GNN) for TSN configuration-space exploration. Rather than replacing formal verification, DeepNC acts as a high-speed pre-verification filter, reserving computationally expensive formal verification only for promising candidates. Extensive evaluations demonstrate that DeepNC significantly improves worst-case delay prediction accuracy over state-of-the-art learning-based methods, increasing the average $R^2$ by 55.8\% and reducing the average MAPE by 65.3\%. More importantly, its high-fidelity regression substantially reduces the number of formal verification calls during configuration-space exploration by 93.25\%, while accelerating NC-based verification by more than two orders of magnitude.

\end{abstract}



\section{Introduction}
Time-Sensitive Networking (TSN)~\cite{IEEE802.1Q} has become the communication backbone for safety-critical systems such as autonomous driving, industrial automation, and aerospace by enabling deterministic communication over Ethernet. However, such determinism is not an inherent property of the TSN standard. Instead, it relies on appropriately configuring diverse scheduling mechanisms (Time-Aware Shaper (TAS)~\cite{ieee8021Qbv}, Credit-Based Shaper (CBS)~\cite{ieee8021Qav}, Cyclic Queuing and Forwarding (CQF)~\cite{8021Qch}, and their hybrid combinations) and performing formal schedulability verification to ensure that the resulting configuration satisfies end-to-end timing requirements. Consequently, practical TSN design relies on an iterative workflow that tightly couples configuration optimization with formal schedulability verification.

Within this iterative design workflow, formal schedulability verification~\cite{Lv20,ZhaoTIE,mohammadpour2023improved,bujosa2024improved} provides the rigorous performance guarantees required for each candidate configuration. Although a single Network Calculus (NC)-based verification typically completes within seconds~\cite{ZhaoTIE}, automated configuration synthesis often requires tens of thousands of repeated verification calls while exploring a large configuration space. Consequently, the computational bottleneck shifts from configuration optimization itself to the repeated execution of formal verification. Existing studies report that schedulability verification can account for over 90\% of the total configuration time in automated TSN design workflows~\cite{reusch2023configuration}.

To alleviate this computational bottleneck, recent studies~\cite{behera2026safety, Mai2019SchedulabilityML, mai2022gnn} have explored learning-based surrogate models that approximate the results of formal verification. Rather than replacing formal methods, these surrogates serve as fast pre-verification modules that rapidly filter unpromising configurations, allowing rigorous analysis to focus only on a much smaller candidate subset. Such a workflow significantly accelerates configuration-space exploration while preserving the correctness guarantees of formal verification. Despite recent progress, existing learning-based pre-verification surrogates remain largely output-driven. They primarily approximate the input-output relationship between network configurations and verification results, without explicitly encoding the analytical reasoning process underlying formal performance analysis methods. Consequently, existing studies have primarily focused on binary schedulability classification~\cite{behera2026safety, Mai2019SchedulabilityML, mai2022gnn}, which is well suited for feasibility screening but provides limited quantitative timing information to guide fine-grained configuration optimization.

To address these limitations, we propose DeepNC, a novel pre-verification surrogate module for TSN configuration-space exploration by structurally integrating NC principles into Graph Neural Networks (GNN)~\cite{scarselli2008graph}. Unlike prior learning-based surrogates that treat formal analysis merely as a source of supervision, DeepNC embeds the reasoning process of the Total Flow Analysis (TFA)-based NC analysis for hybrid TSN/TAS+CBS architectures~\cite{ZhaoTIE} into the graph message-passing mechanism of the neural architecture. As a result, DeepNC closely approximates worst-case delay bound (WCDs) computed by NC formal analysis, providing quantitative timing information for efficient configuration-space exploration and candidate pruning in large-scale TSN networks. Our main contributions are as follows:
\begin{itemize}

\item First, we propose an NC-guided heterogeneous graph representation to encode the analytical semantics of NC. By deconstructing the network into distinct \myconcept{Flow}, \myconcept{Queue}, and \myconcept{Link} nodes, we explicitly map the defining parameters of arrival curves and service curves into graph features, enabling DeepNC to preserve the analytical structure of NC.

\item Second, we develop a message-passing mechanism inspired by the TFA-based NC analysis. It consists of two interleaved phases: a Path-Level Update, which employs Gated Recurrent Units (GRUs) to capture the evolution of arrival curves, and a Port-Level Update, which aggregates flow information to update service curves under resource contention. This incorporates NC domain knowledge into GNN learning, enabling DeepNC to achieve high-fidelity delay regression.

\item Third, extensive test cases, including both synthetic and realistic, demonstrate that DeepNC serves as an accurate and efficient pre-verification surrogate for TSN configuration workflows. Compared with state-of-the-art learning-based methods, DeepNC achieves significantly higher regression fidelity, improving the average $R^2$ by 55.8\% while reducing MAPE by 65.3\%. As a result, DeepNC accelerates NC verification by over two orders of magnitude and reduces formal calls by 93.25\% on average, significantly outperforming SOTA baselines.
\end{itemize}

The rest of the paper is organized as follows: We first review related work in Sec.~\ref{sec:RelatedWork}. Sec.~\ref{sec:TSNSystemModel} presents the system model and NC background. Our proposed pre-verification surrogate module, DeepNC, is detailed in Sec.~\ref{sec:DeepNC}. Extensive evaluations are presented in Sec.~\ref{sec:Experimental Results}, and Sec.~\ref{sec:Conclusion} concludes the paper.

\section{Related Work}
\label{sec:RelatedWork}
The growing demand for large-scale and dynamic TSN has motivated research into rapid network configuration~\cite{boyer2025embedded}. To address the bottleneck of formal verification in configuration-space exploration, existing approaches can be broadly classified into two categories: a priori awareness~\cite{TII, Xie, finzi2024integrating, bujosa2026efficient, ashjaei2017schedulability} and a posteriori verification~\cite{barzegaran2022real,gavriluct2020traffic,berisa2022avb}. The \textit{a priori} awareness paradigm integrates formal verification analysis into the optimization process by deriving specialized analytical or semi-analytical models for specific scheduling policies or optimization objectives~\cite{TII, Xie, finzi2024integrating, bujosa2026efficient, ashjaei2017schedulability}. While computationally efficient, these methods require problem-specific mathematical formulations and are therefore difficult to generalize.

In contrast, the \textit{a posteriori} verification paradigm~\cite{barzegaran2022real,gavriluct2020traffic,berisa2022avb} preserves the conventional optimization-verification workflow while accelerating verification itself through techniques such as incremental verification~\cite{Zhao24Incremental}, though their effectiveness diminishes with an increasing number of concurrent changes. More recently, learning-base surrogate models have been introduced to accelerate formal verification~\cite{behera2026safety, Mai2019SchedulabilityML,mai2022gnn}. Behera et al.~\cite{behera2026safety} propose a method for performing binary schedulability classification for variable-length periodic task sets under multiple uniprocessor scheduling policies, serving as a pre-filter before formal schedulability analysis. In the networking domain, the RouteNet family~\cite{rusek2020routenet, ferriol2022routenet, ferriol2023routenetfermi} and CEA-GNN~\cite{CEA-GNN} predict average network performance. In contrast, Mai et al.~\cite{Mai2019SchedulabilityML,mai2022gnn} formulate Ethernet/TSN verification as binary schedulability classification.

However, all these methods primarily learn an input-output mapping without explicitly modeling the underlying analytical reasoning process. As a result, they are less suitable for high-fidelity WCDs prediction required for refined configuration optimization. In contrast, DeepNC, proposed in this paper, structurally integrates the analytical principles of TFA-based NC analysis into GNN to support high-fidelity worst-case delay regression.

\section{TSN System Model}
\label{sec:TSNSystemModel}
\subsection{System Model}
\label{sec:SystemModel}
\begin{figure}[!t]   
	\centering
	\includegraphics[width=1\linewidth]{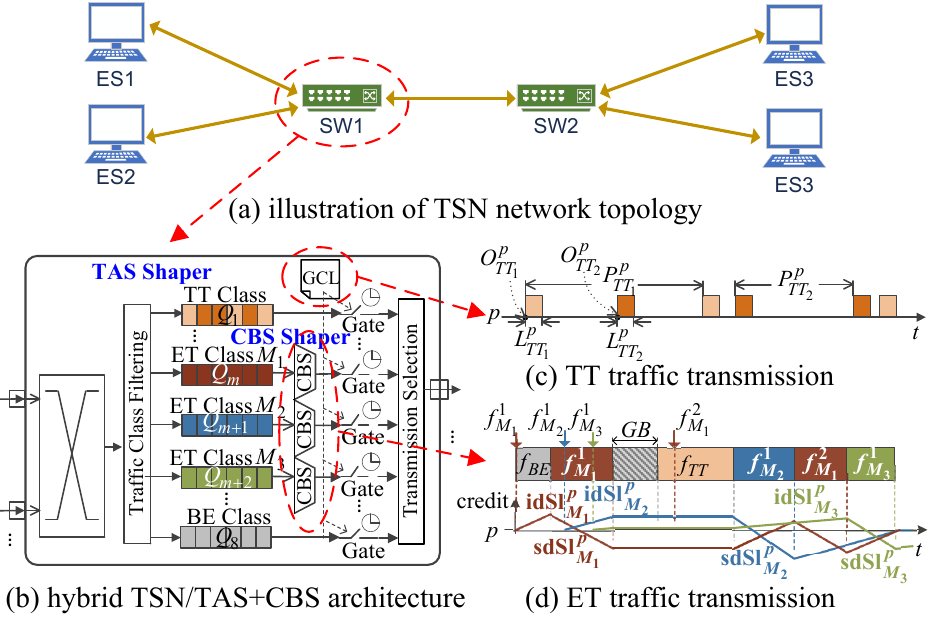}  
	\caption{TSN/TAS+CBS Hybrid Architecture}
    \label{fig:TAS+CBS}
    \vspace{-0.2cm}
\end{figure}

In this work, we consider a TSN with an arbitrary topology, composed of End Systems (ESs) and Switches (SWs) interconnected by full-duplex links of rate $C$. Each ES connects to a switch, and switches connect to other switches or ESs via their egress ports, where there is one-to-one mapping between a link and an output port. Fig.~\ref{fig:TAS+CBS}(a) illustrates a typical example, consisting of four ESs and two SWs, where the double-headed arrows represent physical links.

Messages are transmitted as flows, which contend for link resources at the output ports of nodes (ESs or SWs). In this paper, we adopt a widely used hybrid TSN scheduling architecture combining the Time-Aware Shaper (TAS) and the Credit-Based Shaper (CBS) to provide real-time guarantees for applications with diverse QoS requirements. As shown in Fig.~\ref{fig:TAS+CBS}(b), this architecture supports three traffic classes: Time-Triggered (TT) flows controlled by TAS, Event-Triggered (ET) flows shaped by CBS, and Best-Effort (BE) traffic. Each output port $p$ has eight queues for these classes, each with its own gate that controls frame forwarding.

TAS is used for periodic, hard real-time TT flows, which are given the highest priority. It employs a Gate Control List (GCL) at each output port $p$ to precisely control the open/close times of the queue gate, ensuring deterministic forwarding of TT as illustrated in Fig.~\ref{fig:TAS+CBS}(c). The GCL at a port 
$p$, denoted as a matrix $\textbf{G}^p$, defines the transmission schedule. Following a flow-based scheduling abstraction, we assume each entry $\hvec{\mathbf{g}}_f^{p}$ in $\textbf{G}^p$ corresponds to a specific TT flow, defined by the tuple $\hvec{\mathbf{g}}_f^{p}=[L_{\TT}^p,P_{\TT}^p,O_{\TT}^p]$, for $\forall f\in \mathcal{F}_{\TT}^p$, where $L_{\TT}^p$, $P_{\TT}^p$ and $O_{\TT}^p$ are the frame length, period, and offset of a TT flow $f$, respectively. We assume all GCLs are pre-configured, and their synthesis is outside the scope of this paper.

CBS schedules lower-priority ET flows, which have less stringent timing requirements. It supports $n_{\CBS}$ priority classes, where class $M_i$ has a higher priority than $M_{i+1}$. To prevent interference with TT flows, CBS queue gates are only permitted to open when the TT gate is closed, allowing ET flows to contend for the remaining bandwidth. The CBS uses a credit mechanism for bandwidth management, governed by an idleSlope parameter $\idSl^p_{M_i}$ for each Class $M_i$ at port $p$. 
An ET frame can only be transmitted if its gate is open and the CBS permits it. Each CBS queue for Class $M_i$ maintains a credit value $c^p_{M_i}$, initialized to zero. When the CBS gate is closed, the associated credit is ``frozen''. When the associated CBS gate is open, the credit decreases by sendSlope $\sdSl^p_{M_i}=\idSl^p_{M_i}-C$ during the transmission and increases with idleSlope $\idSl^p_{M_i}$ when
ET frames are waiting to be transmitted,
as shown for example in Fig.~\ref{fig:TAS+CBS}(d) with three CBS classes. For each ET flow $f$, its frame size $L_f$ and minimum inter-arrival time $P_f$ sending from the source ES are known. We denote $\mathcal{F}_{M_i}^p$ as the set of flows of Class $M_i$ at queue $q_{M_i}$ of port $p$. 

The remaining queues are used for BE flows, which have the lowest priority and do not provide real-time guarantees.
\vspace{-0.2cm}
\subsection{Network Calculus Background}
\label{sect:NCBackground}
\vspace{-0.2cm}
Network Calculus (NC), which serves as the theoretical foundation guiding the architecture of DeepNC, is a rigorous mathematical framework for worst-case analysis of queuing systems, employing min-plus algebra to derive deterministic bounds on traffic and service processes~\cite{LeBoudec01, Bouillard18}. 
Total Flow Analysis (TFA), a prominent method within the NC framework, provides a methodology to aggregate per-node results. 
Let $\mathcal{F}_\uparrow$ be the set of non-decreasing functions from non-negative reals $\mathbb{R}^+$ to $\mathbb{R}^+$.

\subsubsection{Per-Node Modeling}
\label{sect:perNodeNC}
NC theory leverages min-plus convolution to provide envelope bounds to abstract the behavior on a single server. This involves characterizing the arriving traffic and the guaranteed service.

The arrival curve $\alpha_f^p(t)$ is used to upper bound the amount of data from a flow $f$ arriving at server $p$ over any time interval. Let $R_f^p(t)\in\mathcal{F}_\uparrow$ denote the cumulative arrival function of the flow, counting the total number of bits that have arrived at a server $p$ by time $t$. We say that $\alpha_f^p(t)$ is an arrival curve for the arrival process $R_f^p(t)\in \mathcal{F}_\uparrow$ of a flow if
\vspace{-0.1cm}
\begin{equation}     \small\label{eq:alpha}
	R_f^p(t)\leq (R_f^p\otimes\alpha_f^p)(t), \ \ \forall t\geq 0,
    \vspace{-0.1cm}
\end{equation} 
where $\otimes$ is the min-plus convolution operator, defined for any $\forall f, g\in \mathcal{F}_\uparrow$ as,
\vspace{-0.1cm}
\begin{equation}     \small\label{g:MinPlusConvolution}
	(f\otimes g)(t)=\inf_{0\leq s\leq t}\{f(t-s)+g(s)\}, \ \ \forall t\in \mathbb{R}^+.
\end{equation}
\vspace{-0.1cm}
A typical arrival curve is the linear ``burst-rate'' envelope,
\begin{equation}     \small\label{eq:burst_rate_f}
  \alpha^{p}_{f}(t)= \sigma_f^{p}+ \rho_f^{p}\cdot t,
\end{equation}
where $\sigma_f^{p}$ and $\rho_f^{p}$ are the maximum burst and long-term rate, respectively. Especially, at the source server $p_0$, $\sigma_f^{p_0}=L_f$ and $\rho_f^{p_0}=L_f/P_f$. Moreover, when multiple flows of the same priority class $M_i$ converge at a server $p$, their combined arrival envelope is modeled by an aggregate arrival curve, $\alpha^{p}_{M_i}(t)$, which is typically the sum of the individual curves,
\begin{equation}     \small\label{eq:arr_Mi}
    \alpha^{p}_{M_i}(t)= \sum_{f\in \mathcal{F}_{M_i}^p}\alpha^{p}_{f}(t).
    \vspace{-0.2cm}
\end{equation}
This aggregate arrival curve is then used to compute the performance bounds for the shared resources, which is fundamental to the TFA method.

The service curve $\beta_{M_i}^p(t)$ defines the lower bound of processing capability of a server $p$ provided to aggregate flows of priority $M_i$. Let $R_{M_i}^{p}(t)\in\mathcal{F}_\uparrow$ and $R_{M_i}^{p*}(t)\in\mathcal{F}_\uparrow$ denote the cumulative arrival and departure functions of these aggregate flows at server $p$, respectively. We say that the server $p$ is to offer a min-plus service curve $\beta_{M_i}^p(t)$ to the aggregate flows of priority $M_i$ if
\begin{equation}     \small\label{eq:beta_conv}
	R_{M_i}^{p*}(t)\geq(R_{M_i}^p\otimes\beta_{M_i}^p)(t).
\end{equation}

Based on these aggregate-level models, the worst-case latency for the entire aggregate traffic of class $M_i$ at server $p$ is upper bounded by the maximum horizontal deviation between the aggregate arrival curve $\alpha_{M_i}^p(t)$ and the service curve $\beta_{M_i}^p(t)$,
\begin{equation}     \small\label{eq:D}	
	D_{M_i}^p=\sup_{t\geq 0}\{\inf\{d \geq 0\ \big|\ \alpha_{M_i}^p(t)\leq \beta_{M_i}^p(t+d)\}\}=D_f^p,
    \vspace{-0.1cm}
\end{equation}
which is also the WCDs for every individual flow $f$ within that aggregate according to the TFA method~\cite{ZhaoTIE, Bouillard18}.

\subsubsection{Network Modeling}
\label{sec:NetworkNC}
The core principle of TFA is to derive an end-to-end delay bound by summing the per-server delay bounds along the flow path. As stated previously, computing the latency bound at any given server $p$ requires knowledge of the input arrival curve of each flow at that server. 
The arrival curve $\alpha_f^{p_0}(t)$ at the source node $p_0$ is directly determined by the specified flow parameters, such as its maximum frame size and minimum frame interval. For any subsequent server $p$ downstream, however, the arrival curve for a flow is altered by the service it received from upstream nodes.

In NC, this evolution is typically captured by computing the output arrival curve $\alpha_f^{p*}(t)$, which upper-bounds the data amount of a flow departing from server $p$ over any time interval,
\begin{equation}     \small\label{eq:alpha_out}
	\alpha_f^{*p}(t)=(\alpha_f^p \oslash \delta_{D_f^p}^p)(t)=\alpha_f^{p+1}(t),
\end{equation}
\vspace{-0.1cm}
where $\delta_{D_f^p}^p(t)$ is the pure-delay function~\cite{LeBoudec01} which equals to 0 if $t\leq D_f^p$ and $+\infty$ otherwise, and $\oslash$ is the min-plus deconvolution operator, defined for any $\forall f, g\in \mathcal{F}_\uparrow$ as,
\begin{equation}     \small\label{g:MinPlusDeconvolution}
	(f\oslash g)(t)=\sup_{s\geq0}\{f(t+s)-g(s)\}, \ \ \forall t\in \mathbb{R}^+.
    \vspace{-0.1cm}
\end{equation}
Note that it in turn serves as the input arrival curve $\alpha_f^{p+1}(t)$ of the flow $f$ for the subsequent hop $p+1$\footnote{In TFA application~\cite{Bouillard18,ZhaoTIE}, burstiness often propagates per-flow because frequent aggregating and splitting make end-to-end aggregate propagation infeasible. For persistent aggregates, pessimism is commonly mitigated via shaping curves.}.

\begin{figure}[!t]
	\centering	\includegraphics[width=1\linewidth]{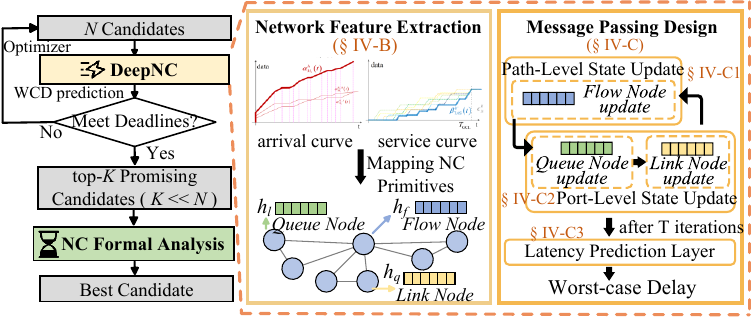}  
	\caption{Overall Framework.}
	\label{fig:overall_framework}
\end{figure}

Subsequently, the end-to-end WCDs for a specific flow $f$ is derived by summing these per-server delay bounds along its route $r_f$,
\begin{equation}     \small\label{g:dE2E}
	D_{\text{E2E},f}=\sum_{p\in r_f} D_f^p.
\end{equation}

\section{DeepNC: A NC Theory-Guided GNN Architecture}
\label{sec:DeepNC}
\subsection{Overall Framework}
\label{sec:OverallFramework}
Our proposed model, DeepNC, is a purpose-designed GNN that structurally integrates the analytical principles of TFA-based NC for hybrid TSN/TAS+CBS networks~\cite{ZhaoTIE}. The overall workflow incorporating DeepNC is illustrated in Fig.~\ref{fig:overall_framework}. During configuration-space exploration, DeepNC does not replace formal verification. Instead, we adopt a filter-then-verify paradigm in which only the top-$K$ configurations predicted by DeepNC are subjected to NC formal analysis, thereby preserving the formal guarantees provided by NC. Additionally, because DeepNC provides high-fidelity delay prediction with substantially lower computational cost than NC analysis, it effectively alleviates the verification bottleneck in large-scale TSN configuration. It is worth noting that since DeepNC is designed to accelerate formal verification rather than replace it, the surrogate is expected to approximate the analytical results of NC as close as possible rather than the exact worst-case network behavior. Therefore, the conservativeness of NC with respect to the actual network behavior does not affect the objective of DeepNC.

The DeepNC framework is implemented in two main stages, as illustrated in the right panel of Fig.~\ref{fig:overall_framework}: Network Feature Extraction (Sec.~\ref{sec:FeatueExtraction}) and Message Passing Design (Sec.~\ref{sec:MessagePassing}). The first stage, Network Feature Extraction, bridges the representation gap by translating the curve functions of NC into a GNN-compatible vector format. To achieve this, we construct a heterogeneous graph with \myconcept{Flow}, \myconcept{Queue}, and \myconcept{Link} nodes from the arrival curve (Sec.~\ref{sec:ACFeatueExtraction}) and the service curve (Sec.~\ref{sec:SCFeatueExtraction}) perspectives. Simultaneously, the key parameters defining these curves are extracted to initialize the feature vectors of their respective nodes. The second stage, Message Passing Design, is engineered to capture the complex dependencies inherent in NC formal analysis. Through customized message-passing steps, the hidden states of the graph nodes are progressively updated to emulate the evolution of arrival curves (Sec.~\ref{sec:MessagePassingPathStateUpdate}) and the impact of inter-queue contention on service curves (Sec.~\ref{sec:MessagePassingPortStateUpdate}). Finally, a readout layer (Sec.~\ref{sec:LatencyPredictionLayer}) utilizes the final \myconcept{Queue Node} states to predict the per-server and end-to-end WCDs for each ET flow. The predicted delay bounds are subsequently used to filter candidate configurations as described in Algo.~\ref{algorithm2}.
\vspace{-0.1cm}
\subsection{Network Feature Extraction: Mapping NC Primitives}
\label{sec:FeatueExtraction}
\vspace{-0.1cm}
The first stage in realizing our DeepNC framework is to bridge the fundamental data representation gap between NC and GNNs. NC operates on curve functions, whereas GNNs require fixed-size feature vectors as input. This section details our methodology for this critical ``function-to-vector'' mapping. Specifically, our approach is to, drawing directly from the foundational concepts of NC (i.e., the arrival and service curves), define a set of graph nodes and construct their corresponding initial feature vectors.

\subsubsection{From the Arrival Curve Perspective}
\label{sec:ACFeatueExtraction}
To represent arrival envelope of each flow in the GNN, we define a \myconcept{Flow Node} as a surrogate for the arrival curve, instantiating one such node per ET flow.

The crucial question, however, is how to define the initial features ($\hvec{\mathbf{x}}_f$) for these \myconcept{Flow Node} instances. As established in our NC background (Section~\ref{sec:NetworkNC}), for an ET flow in TSN networks, the burst $\sigma_f^{p_0}$ and long-term rate $\rho_f^{p_0}$ correspond directly to the frame size $L_f$ and the ratio $L_f/P_f$. 
Therefore, the initial feature vector for each \myconcept{Flow Node} instance is defined based on its source characteristics. This vector contains the parameters that fully define the source arrival curve $\alpha^{p_0}_{f}(t)$. The \myconcept{Flow Node} feature vector is formally defined as follows.

\begin{definition}\label{definition:FlowNode}
	(\myconcept{Flow Node} Feature Extraction) A \myconcept{Flow node} instance is designed to abstract the arrival curve of an individual ET flow $f$, with its initial features captured from source ES:
	\begin{equation}     \small\label{eq:xf}
        \hvec{\mathbf{x}}_f^{p_0}=[L_f,P_f],
        \vspace{-0.1cm}
    \end{equation}
where $L_f$ and $P_f$ are respectively the frame size and minimum time interval between two consecutive frames of $f$.
\end{definition}

We use raw $[L_f,P_f]$ rather than precomputed long-term rate $\rho_f^{p_0}$ to reduce preprocessing on large-scale graphs and enable the GNN to implicitly learn the rate during training.

\subsubsection{From the Service Curve Perspective}
\label{sec:SCFeatueExtraction}
The goal is to represent the service guarantees for specific flows at each server within the GNN. In TSN networks featuring a hybrid TSN/TAS+CBS architecture, the service curve $\beta^p_{M_i}(t)$ for the aggregate flows of CBS Class $M_i$ at an egress port $p$ is explicitly given by~\cite{ZhaoTIE},
\vspace{-0.2cm}
\begin{equation}     \small\label{eq:beta_mi}
    \beta^p_{M_i}(t)=\idSl^p_{M_i}\cdot\left[\left(t-\frac{\alpha^p_{\TT}(t)}{C}-\frac{c_{M_i}^{p,\max}}{\idSl^p_{M_i}}\right)\right]^+_\uparrow,
    \vspace{-0.1cm}
\end{equation}
where $\alpha^p_{\TT}(t)$ is the aggregate arrival curve of higher-priority TT traffic derived from the GCL matrix $\textbf{G}^p$, $[f(t)]^+_\uparrow = \max_{0\leq s\leq t}\{f(s),0\}$, and $c_{M_i}^{p,\max}$ is the credit upper bound, which is influenced by contention from other CBS classes.

All traffic classes share port $p$ and this contention is reflected in the service curve in Eq.~(\ref{eq:beta_mi}). Its parameters fall into two groups: (i) queue-specific: idle slope $\idSl_{M_j}^{p}$ and credit bound $c_{M_i}^{p,\max}$. (ii) port-shared: physical link capacity $C$ and TT interference $\alpha^p_{\TT}(t)$. 
This parametric duality motivates us to represent the CBS server by two dedicated graph node types: (a) We term \myconcept{Queue Node} to represent the queue-specific parameters for each CBS Class $M_i$.
(b) We term \myconcept{Link Node} to represent the port-shared parameters. 
This design choice is crucial: it avoids redundant modeling of shared parameters (e.g., repeating $C$ and GCL information in every queue node at the same port) and allows the GNN to learn the interactions between queues and their shared link through message passing, as detailed in Sec.~\ref{sec:MessagePassing}. Thus, we instantiate one \myconcept{Queue Node} for each CBS class at a port, and a single \myconcept{Link Node} for the port itself.

The initial features $\hvec{\mathbf{x}}_{q_{M_i}}^p$ for a \myconcept{Queue Node} must capture parameters that influence the credit upper bound $c_{M_i}^{p,\max}$, which is dependent on other competing CBS priority classes. To address this, we introduce a one-hot vector $\hvec{\mathbf{e}}_{M_i}^p$ to encode the priority information, providing a basis for the GNN to subsequently learn the interactive effects from these competing priorities during the message passing phase (Sec.~\ref{sec:MessagePassing}). The \myconcept{Queue Node} feature vector is formally defined as follows.

\begin{definition}\label{definition:QueueNode}
	(\myconcept{Queue Node} Feature Extraction) A \myconcept{Queue Node} instance is designed to abstract the queue-specific part of service curve for a CBS Class $M_i$ at port $p$, with its initial features captured as:
	\begin{equation}     \small\label{eq:xqmi}    \hvec{\mathbf{x}}_{q_{M_i}}^p=\left[\idSl^p_{M_i}, \hvec{\mathbf{e}}_{M_i}^p\right],
    \end{equation}
    Here, $\hvec{\mathbf{e}}_{M_i}^p=[0,...1,...0]$ is a one-hot vector encoding the CBS priority, where the bit at the index for Class $M_i$ is set to 1 and 0 otherwise. This vector provides the structural basis for learning priority contention effects via message passing.
\end{definition}

Similarly, the initial features $\hvec{\mathbf{x}}_l^p$ for a \myconcept{Link Node} must capture the aggregate TT arrival curve $\alpha^p_{\TT}(t)$. As detailed in~\cite{ZhaoTIE}, $\alpha^p_{\TT}(t)$ is a deterministic, hyperperiodic step function fully determined by the GCL\footnote{As discussed in Sec.~\ref{sec:OverallFramework}, DeepNC is designed to approximate the analytical results of the TFA-based NC formulation in~\cite{ZhaoTIE} as faithfully as possible. Therefore, we adopt the same TT step function model as in~\cite{ZhaoTIE}.}. However, computing the parameters of its tightest envelope is a computationally intensive process. To avoid this high feature extraction cost, we again adopt a strategy of using the raw, fundamental parameters. Instead of pre-calculating the envelope $\alpha^p_{\TT}(t)$, we directly extract the information from the columns of the GCL matrix $\textbf{G}^p$. Specifically, we form a frame size vector $\hvec{\mathbf{g}}_L^p$, a period vector $\hvec{\mathbf{g}}_P^p$, and an offset vector $\hvec{\mathbf{g}}_O^p$. These vectors serve to implicitly capture the information of $\alpha^p_{\TT}(t)$ without explicit computation. The \myconcept{Link Node} feature vector is formally defined as follows.

\begin{definition}\label{definition:LinkNode}
	(\myconcept{Link Node} Feature Extraction) A \myconcept{Link Node} instance is designed to abstract the port-shared part of the service curve at an egress port $p$, with its initial features captured as:
	\begin{equation}     \small\label{eq:xl}
        \hvec{\mathbf{x}}_l^p= \big[\hvec{\mathbf{g}}_L^p,\hvec{\mathbf{g}}_P^p,\hvec{\mathbf{g}}_O^p, C \big],
    \end{equation}
	where $\hvec{\mathbf{g}}_L^p$, $\hvec{\mathbf{g}}_P^p$, and $\hvec{\mathbf{g}}_O^p$ are the column vectors derived from the GCL matrix $\textbf{G}^p$, representing the frame sizes, periods and offsets of TT flows, respectively.
\end{definition}

We have now established the three types of graph nodes in our DeepNC framework, i.e., \myconcept{Flow Node}, \myconcept{Queue Node}, \myconcept{Link Node}, and their initial feature vectors ($\hvec{\mathbf{x}}_f^{p_0}$, $\hvec{\mathbf{x}}_{q_{M_i}}^p$, $\hvec{\mathbf{x}}_l^p$). These features are grounded in NC theory, bridging the ``function-to-vector'' gap. However, the graph constructed is heterogeneous: the initial feature vectors for each node type reside in disparate semantic spaces, possessing inconsistent dimensions and different physical units (e.g., bit vs. s vs. bps). To ensure mathematical and semantic compatibility for subsequent operations, we employ distinct multi-layer perceptrons (MLPs)~\cite{pal1992multilayer} to project these raw features into a unified latent space. This mapping transforms the initial feature vectors ($\hvec{\mathbf{x}}_f^{p_0}$, $\hvec{\mathbf{x}}_{q_{M_i}}^p$, $\hvec{\mathbf{x}}_l^p$) into initial hidden states ($\hvec{\mathbf{h}}_{f}^{t_0, p_0}$, $\hvec{\mathbf{h}}_{q_{M_i}}^{t_0,p}$, $\hvec{\mathbf{h}}_{l}^{t_0,p}$) of a unified dimension, 

\begin{equation}     \small
	\begin{cases}
		\hvec{\mathbf{h}}_{f}^{t_0, p_0} = \text{MLP}_1(\hvec{\mathbf{x}}_f^{p_0})  \\
        \hvec{\mathbf{h}}_{q_{M_i}}^{t_0,p} = \text{MLP}_2(\hvec{\mathbf{x}}_{q_{M_i}}^p)  \\
        \hvec{\mathbf{h}}_{l}^{t_0,p} = \text{MLP}_3(\hvec{\mathbf{x}}_l^p) .
	\end{cases}
\end{equation}
These hidden states then serve as the input for the first iteration $t_0$ of message passing.

\subsection{Message Passing Design}
\label{sec:MessagePassing}
The next stage is to design a message passing mechanism that emulates the NC calculation trajectory. As described in Sec.~\ref{sect:NCBackground}, the worst-case end-to-end delay bound results from a multi-step process involving (a) arrival curve evolution along the path and (b) service curve affected under inter-queue contention of different priorities. Our message passing layers are designed to simulate these path-dependent and contention-aware processes by directly mirroring the core NC operations. The mechanism iteratively updates the hidden states ($\hvec{\mathbf{h}}$) of nodes, which serve as vector-based representations for evolving NC curves. By aligning the GNN with the mathematical principles of NC, the model effectively captures the underlying physics of network worst-case end-to-end delay accumulation.

In the following, we detail how DeepNC simulates the NC calculation through message passing among its \myconcept{Flow}, \myconcept{Queue}, and \myconcept{Link} nodes to predict the WCDs, as outlined in Algo.~\ref{algorithm 1}. The message passing is designed as an iterative process over $T$ iterations (indexed by $t=t_0, ..., t_{T-1}$). Each iteration consists of two main phases: A Path-Level State Update phase, which tracks the evolution of arrival curves along each flow path. A Port-Level State Update phase, which refines the service curve representations by incorporating aggregated traffic load and inter-queue contention information. These two phases are implemented as three types of state updates for the \myconcept{Flow}, \myconcept{Queue}, and \myconcept{Link} nodes.

\subsubsection{Path-Level State Update (Algo.~\ref{algorithm 1} Lines 2-5)}
\label{sec:MessagePassingPathStateUpdate}
For a given flow $f$, its hidden state $\hvec{\mathbf{h}}_{f}^{t,p}$ represents its arrival curve at egress port $p$ along its path. The initial hidden state $\hvec{\mathbf{h}}_{f}^{t_0,p_0}$ is defined for the source port and remains fixed. This phase simulates the evolution of the arrival curve of each flow along its multi-hop path. In NC theory, the output arrival curve $\alpha_f^{*p}(t)$ from port $p$ becomes the input arrival curve $\alpha_f^{p+1}(t)$ for the next hop $p+1$. As given by Eq.~(\ref{eq:alpha_out}), $\alpha_f^{*p}(t)$ is the min-plus deconvolution of $\alpha_f^p(t)$ with the pure-delay function $\delta_{D_f^p}(t)$. Therefore, the GNN needs representations for both the input arrival curve (provided by $\hvec{\mathbf{h}}_{f}^{t,p}$) and the pure-delay function, which is encoded in \myconcept{Queue Node} state $\hvec{\mathbf{h}}_{q_{M_i}}^{t,p}$ for CBS Class $M_i$\footnote{Following the TFA methodology~\cite{ZhaoTIE, Bouillard18}, the delay bound derived for aggregate flows of the same class $M_i$ applies to every flow within that aggregate, as shown by Eq.~(\ref{eq:D}).} after port-level state updates (Sec.~\ref{sec:MessagePassingPortStateUpdate}). Note that the \myconcept{Queue Node} state $\hvec{\mathbf{h}}_{q_{M_i}}^{t,p}$ encodes latency information only after the message-passing iterations ($t>t_0$).

With the hidden state representations of the input arrival curve $\alpha_f^p(t)$ and the pure-delay function $\delta_{D_f^p}(t)$ available, we can now formally define the state update for the \myconcept{Flow Node}. We employ a Gated Recurrent Unit (GRU)~\cite{ferriol2023routenetfermi}, a type of recurrent neural network adept at modeling stateful, sequential transformations, which aligns well with the hop-by-hop nature of the NC process of transforming an arrival curve.
\begin{definition}[\myconcept{Flow Node} State Update]
    A GRU cell is applied at each hop $p$. Its input consists of the current \myconcept{Flow Node} hidden state $\hvec{\mathbf{h}}_{f}^{t,p}$ and the \myconcept{Queue Node} hidden state $\hvec{\mathbf{h}}_{q_{M_i}}^{t,p}$, and its output is the updated hidden state $\hvec{\mathbf{h}}_{f}^{t,p+1}$, which represents the output arrival curve of flow $f$ at port $p$ and serves as the input arrival curve for the next hop $p+1$:
    \begin{equation}     \small\label{eq:mft_new}
        \hvec{\mathbf{h}}_{f}^{t,p+1} = \text{GRU} \left( \hvec{\mathbf{h}}_{f}^{t,p}, \hvec{\mathbf{h}}_{q_{M_i}}^{t,p} \right),
    \end{equation}
\end{definition}

It is worth noting that as message-passing iterations ($t>t_0$) proceed, $\hvec{\mathbf{h}}_{f}^{t,p+1}$ is progressively refined as $\hvec{\mathbf{h}}_{q_{M_i}}^{t,p}$ aggregates increasingly richer contextual information (Eq.~(\ref{eq:QueueNodeUpdate})), thereby better representing the corresponding NC arrival curve. Additionally, since the source arrival curve is deterministic, $\hvec{\mathbf{h}}_{f}^{t,p_0}$ at source ES $p_0$ remains fixed: $\hvec{\mathbf{h}}_{f}^{t+1,p_0}=\hvec{\mathbf{h}}_{f}^{t,p_0}=\hvec{\mathbf{h}}_{f}^{t_0,p_0}$.

\subsubsection{Port-Level State Update (Algo.~\ref{algorithm 1} Lines 6-11)}
\label{sec:MessagePassingPortStateUpdate}
After the path-level state update phase, we proceed to the port-level state update phase. This phase updates the hidden states of the \myconcept{Queue} and \myconcept{Link} nodes, preparing them for the next iteration $t+1$ to refine the NC reasoning process.

The purpose of updating the \myconcept{Queue Node} state $\hvec{\mathbf{h}}^{t,p}_{q_{M_i}}$ is to create an aggregated representation of both the arrival curve and service curve for CBS Class $M_i$ at port $p$. This aggregated representation is then used to ultimately predict the delay bound $D_{M_i}^p$ of aggregate flows of CBS Class $M_i$ at port $p$, as according to Eq.~(\ref{eq:D}), which is jointly determined by the aggregate arrival curve $\alpha_{M_i}^p(t)$ and the service curve $\beta_{M_i}^p(t)$.

We first construct a vector representation $\hvec{\bm{\beta}}^{t,p}_{M_i}$ of the service curve $\beta_{M_i}^p(t)$ for the current iteration $t$. As shown in Eq.~(\ref{eq:beta_mi}), the service curve depends on both queue-specific parameters (e.g., idle slope) and port-shared parameters (e.g., link capacity and TT traffic interference). As presented in Sec.~\ref{sec:FeatueExtraction}, these two types of information are encoded in the \myconcept{Queue Node} state $\hvec{\mathbf{h}}_{q_{M_i}}^{t,p}$ and the \myconcept{Link Node} state $\hvec{\mathbf{h}}_{l}^{t,p}$, respectively. Therefore, we fuse them through an MLP to obtain the service curve representation:
\begin{equation}     \small\label{eq:vt_new}
    \hvec{\bm{\beta}}^{t,p}_{M_i} = \text{MLP}_{\text{S}} \left( \left[ \hvec{\mathbf{h}}_{q_{M_i}}^{t,p} \parallel \hvec{\mathbf{h}}_{l}^{t,p} \right] \right).
\end{equation}

Next, we compute an intermediate hidden state $\hvec{\bm{\alpha}}_{M_i}^{t,p}$ to represent the aggregate arrival curve $\alpha_{M_i}^p(t)$ as given by Eq.~(\ref{eq:arr_Mi}). This is achieved by summing the hidden states of all \myconcept{Flow Node} instances corresponding to CBS Class $M_i$ passing through port $p$,
\begin{equation}     \small\label{eq:amit}
    \hvec{\bm{\alpha}}_{M_i}^{t,p} = \sum_{f\in \mathcal{F}_{M_i}^p}\hvec{\mathbf{h}}_{f}^{t,p}. 
\end{equation}

With both $\hvec{\bm{\alpha}}_{M_i}^{t,p}$ and $\hvec{\bm{\beta}}_{M_i}^{t,p}$ available, we can formally define the state update for the \myconcept{Queue Node}.

\begin{algorithm}[t!]
	\small
	\SetAlFnt{\small}
	\SetAlCapFnt{\small}
	\SetAlCapSkip{0.3em}
	\SetCommentSty{textit}
	
	\caption{DeepNC Pseudocode}
	\label{algorithm 1}
	
	\KwIn{Flow set $\mathcal{F}$, Queue set $\mathcal{Q}$, Link set $\mathcal{L}$, initial embeddings $\hvec{\mathbf{h}}_{f}^{t_0, p_0}$, $\hvec{\mathbf{h}}_{q_{M_i}}^{t_0,p}$, $\hvec{\mathbf{h}}_{l}^{t_0,p}$, number of iterations $T$}
	\KwOut{Predicted end-to-end worst-case delays $\hat{y}_{D_{\text{E2E},f}}$}
	
	\For{$t \leftarrow t_0$ \KwTo $t_{T-1}$}{
		\ForEach{$f \in \mathcal{F}$ }{
			\tcp{Path-Level State Update (Sec.~\ref{sec:MessagePassingPathStateUpdate})}
			$\hvec{\mathbf{h}}_f^{t,p_0} \gets \hvec{\mathbf{h}}_f^{t_0,p_0}$ \;
			\ForEach{$p \in \mathcal{P}_f$ }{
				$\hvec{\mathbf{h}}_f^{t,p+1} \gets \text{GRU} \left( \hvec{\mathbf{h}}_f^{t,p}, \hvec{\mathbf{h}}_{q_{M_i}}^{t,p} \right)$;  \tcp*[f]{Eq.~(\ref{eq:mft_new})}
			}
		}
		\ForEach{$p \in \mathcal{L}$ }{
			\tcp{Port-Level State Update (Sec.~\ref{sec:MessagePassingPortStateUpdate})}
			\ForEach{$q_{M_i} \in \mathcal{Q}_p$}{
				$\hvec{\bm{\beta}}^{t,p}_{M_i} \gets \text{MLP}_{\text{S}} \left( \left[ \hvec{\mathbf{h}}_{q_{M_i}}^{t,p} \parallel \hvec{\mathbf{h}}_{l}^{t,p} \right] \right)$; \tcp*[f]{Eq.~(\ref{eq:vt_new})}
				
				$\hvec{\bm{\alpha}}_{M_i}^{t,p} \gets \sum_{f \in \mathcal{F}_{M_i}^p} \hvec{\mathbf{h}}_f^{t,p}$; \tcp*[f]{Eq.~(\ref{eq:amit})}
				
				$\hvec{\mathbf{h}}_{q_{M_i}}^{t+1,p} \gets \text{MLP}_{\text{Q}} \left( \left[ \hvec{\bm{\alpha}}_{M_i}^{t,p} \parallel \hvec{\bm{\beta}}_{M_i}^{t,p} \right] \right)$; \tcp*[f]{Eq.~(\ref{eq:QueueNodeUpdate})}
				
			}
			$\hvec{\mathbf{h}}_l^{t+1,p} \gets \text{MLP}_{\text{L}} \left( \left[ \hvec{\mathbf{h}}_l^{t,p} \parallel \sum_{j=1}^{n_{\text{CBS}}} \hvec{\mathbf{h}}_{q_{M_j}}^{t+1,p} \right] \right)$\;  \tcp*[f]{Eq.~(\ref{eq:hl})}
		}
	}
	\ForEach{$f \in \mathcal{F}$ }{
		\tcp{Latency Prediction Layer (Sec.~\ref{sec:LatencyPredictionLayer})}
		$ \hat{y}_{D_{\text{E2E},f}} \gets \sum_{p \in \mathcal{P}_f} \text{MLP}_{\text{Readout}} \left( \hvec{\mathbf{h}}_{q_{M_i}}^{t_T,p} \right)$\; \tcp*[f]{Eqs.~\eqref{eq:yqm}, \eqref{eq:yf}}
	}
\end{algorithm}

\begin{figure}[!t]
	\centering
	\includegraphics[width=1\linewidth]{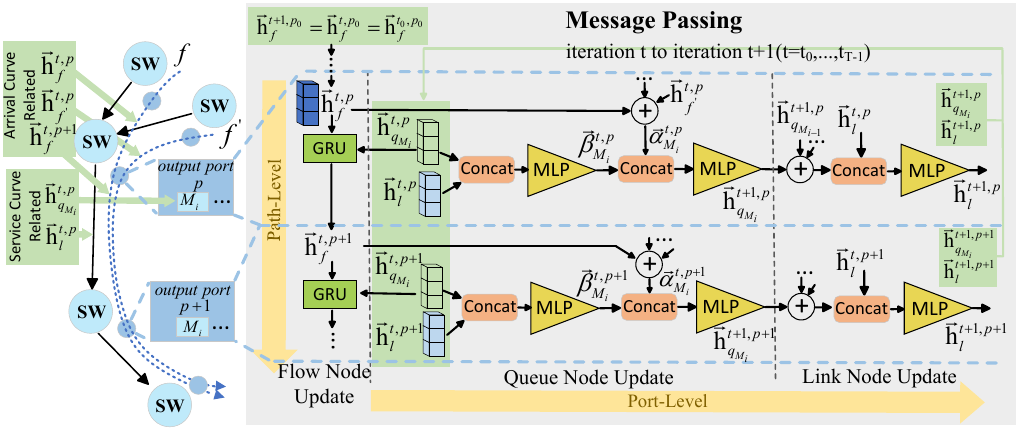} 
	\caption{Message Passing of DeepNC.}
    \vspace{-0.2cm}
	\label{fig:messagepassing}
\end{figure}

\begin{definition}\label{definition:QueueNodeStateUpdate}
	(\myconcept{Queue Node} State Update) The hidden state of a \myconcept{Queue Node} for CBS Class $M_i$ at port $p$ is updated for the next iteration $t+1$. The update is performed by an MLP that takes the concatenation of the aggregate arrival hidden state and the service hidden state from the current iteration $t$ as its input:
\begin{equation}
    \small
    \label{eq:QueueNodeUpdate}
    \hvec{\mathbf{h}}_{q_{M_i}}^{t+1,p} = \text{MLP}_\text{Q}\left(\left[\hvec{\bm{\alpha}}_{M_i}^{t,p} \parallel \hvec{\bm{\beta}}_{M_i}^{t,p}\right]\right). 
\end{equation}
It is worth noting the evolution of this state. At the initial iteration ($t_0$), the hidden state $\hvec{\mathbf{h}}_{q_{M_i}}^{t_0,p}$ primarily encodes its own service parameters (e.g., idle slope $idSl_{M_i}^p$ and priority index). However, in subsequent iterations ($t>t_0$), the \myconcept{Queue Node} state $\hvec{\mathbf{h}}_{q_{M_i}}^{t,p}$ will have additionally incorporated the delay bound information derived from the interaction between the aggregate arrival curve and service curve.
\end{definition}
The purpose of updating the \myconcept{Link Node} state $\hvec{\mathbf{h}}_{l}^{t,p}$ is to enrich the service hidden state $\hvec{\bm{\beta}}_{M_i}^{t+1,p}$ of the next iteration with information about inter-queue priority contention. This is achieved by updating the \myconcept{Link Node} state using the aggregated information from all \myconcept{Queue Nodes} $\hvec{\mathbf{h}}_{q_{M_j}}^{t+1,p}$ ($j\!\!\in \!\![1,n_{\CBS}]$) at the same port. Consequently, when the \myconcept{Link} state $\hvec{\mathbf{h}}_{l}^{t+1,p}$ is used in the next iteration to synthesize the service hidden state via Eq.~(\ref{eq:vt_new}), the new $\hvec{\bm{\beta}}_{M_i}^{t+1,p}$ inherently reflects the contention from all competing CBS classes.
\begin{definition}\label{definition:LinkNodeStateUpdate}
	(\myconcept{Link Node} State Update) The hidden state of a \myconcept{Link Node} at port $p$ is updated for the next iteration $t+1$. The update is performed by an MLP that takes the concatenation of its own state from the current iteration, $\hvec{\mathbf{h}}_{l}^{t,p}$, and the sum of the newly updated states of all \myconcept{Queue Nodes} at that port, $\hvec{\mathbf{h}}_{q_{M_j}}^{t+1,p}, j\in [1,n_{\CBS}]$, as its input:
    \begin{equation}     \small\label{eq:hl}
        \hvec{\mathbf{h}}_{l}^{t+1,p} = \text{MLP}_\text{L}\left(\left[\hvec{\mathbf{h}}_{l}^{t,p} \parallel \sum_{j \in [1,n_{\CBS}]} \hvec{\mathbf{h}}_{q_{M_j}}^{t+1,p}\right]\right).
    \end{equation}

It is worth noting that at the initial iteration ($t_0$), the state $\hvec{\mathbf{h}}^{t_0,p}_l$ of each \myconcept{Link Node} is constructed purely from its own initial features, and thus lacks any information regarding inter-queue priority contention. The message passing described above is shown in Fig.~\ref{fig:messagepassing}.
\end{definition}

\subsubsection{Latency Prediction Layer (Algo.~\ref{algorithm 1} Lines 12-13)}
\label{sec:LatencyPredictionLayer}
After $T$ iterations of message passing, the final hidden state $\hvec{\mathbf{h}}_{q_{M_i}}^{t_{T},p}$ of the \myconcept{Queue Node} has comprehensively encoded the information of both the aggregate arrival curve $\alpha_{M_i}^p(t)$ and the service curve $\beta_{M_i}^p(t)$.
At this point, we can leverage this rich representation to predict the per-server delay bound. We employ a readout MLP layer to emulate the computation from the curves ($\alpha_{M_i}^p(t)$, $\beta_{M_i}^p(t)$) to the delay bound (Eq.~(\ref{eq:D})). The input to this MLP is the final \myconcept{Queue Node} state:
\begin{equation}     \small\label{eq:yqm}
    \hat{y}_{D_{M_i}^p} = \text{MLP}_{\text{Readout}} \left( \hvec{\mathbf{h}}_{q_{M_i}}^{t_T,p} \right)=\hat{y}_{D_f^p}.
\end{equation}
As according to Eq.~\eqref{eq:D}, the delay bound of an aggregate of class $M_i$ flows applies to any individual flow, so we have $D_{M_i}^p=D_f^p$, and thus also the prediction $\hat{y}_{D_{M_i}^p}=\hat{y}_{D_f^p}$. Finally, according to Eq.~\eqref{g:dE2E}, the end-to-end predicted delay bound for flow $f$ is the sum of the predicted per-hop bounds along its path:
\begin{equation}     \small\label{eq:yf}
    \hat{y}_{D_{\text{E2E},f}} = \sum_{p \in \mathcal{P}_f} \hat{y}_{D_f^p}.
\end{equation}

\begin{algorithm}[t!]
	\small\SetAlFnt{\small}\SetAlCapFnt{\small}\SetAlCapSkip{0.3em}\setstretch{1.0}\SetCommentSty{textit}
	\caption{Two-Stage Configuration Verification Based on DeepNC}
	\label{algorithm2}
	\KwIn{$\mathcal{C}_{total}$ (candidate configuration set), $\mathcal{T}_{dead,f}$ (per-flow deadlines), $K$ (number of top candidates to retain)}
	\KwOut{$C^*$ (optimal configuration verified by NC formal analysis)}
	
	$\mathcal{C}_{feasible} \gets \emptyset$\;
	
	\ForEach{$C \in \mathcal{C}_{total}$}{
		$\{ \hat{y}_{D_{\text{E2E},f}} \} \gets \text{DeepNC}(C)$;  \tcp*[f]{DeepNC (fast)}
		
		\If{$\max(\{ \hat{y}_{D_{\text{E2E},f}} - \mathcal{T}_{dead,f} \}) \le 0$}{
			$\mathcal{C}_{feasible} \gets \mathcal{C}_{feasible} \cup \{C\}$\;
		}
		\Else{
			$C' \gets \text{TuningOpt}(C, \{ \hat{y}_{D_{\text{E2E},f}} \})$\;
			$\mathcal{C}_{total} \gets \mathcal{C}_{total} \cup \{C'\}$\;
		}
	}
	$\mathcal{C}_{topk} \gets \text{TopK}(\mathcal{C}_{feasible}, K)$\;
    
	$C^* \gets \arg \mathrm{opt}_{C \in \mathcal{C}_{topk}} \text{metric}(\text{NC\_Tool}(C))$\;
  		\tcp*[f]{Formal NC (safe but slow)}
        
	\Return{$C^*$}\;
\end{algorithm}
Based on the above design, DeepNC provides end-to-end WCD predictions for all ET flows under a given configuration with extremely low inference latency, making it an ideal pre-screening tool for large-scale configuration verification. Algo.~\ref{algorithm2} presents a two-stage DeepNC-based pre-verification process with integrated configuration tuning. In the first stage (Lines~1-9), DeepNC rapidly evaluates all candidates. Configurations meeting deadlines are added directly to the feasible set $\mathcal{C}_{feasible}$, while the others are passed to a designer-defined tuning strategy $\text{TuningOpt}(\cdot)$. Unlike binary schedulability classification~\cite{Mai2019SchedulabilityML,mai2022gnn}, designers can customize the tuning strategy (e.g., adjusting bandwidth allocation or routing paths) based on the deviation between the predicted WCDs and their deadlines. This may convert infeasible configurations into potentially feasible ones while enriching the pool of high-quality candidates. Each tuned configuration $C'$ is then reinserted into the candidate set $\mathcal{C}_{total}$ for subsequent evaluation. After traversal, the top $K$ configurations in $\mathcal{C}_{feasible}$ are selected according to the designer-specified performance metric to form $\mathcal{C}_{topk}$. In the second stage (Lines~10-11), only these $K$ candidates are evaluated by the formal NC analysis tool (safe but slow), and the configuration that optimizes the designer-specified performance metric is selected as the final configuration $C^*$.

\vspace{-0.2cm}

\section{Experimental Results}
\label{sec:Experimental Results}
\vspace{-0.1cm}
This section systematically evaluates DeepNC across estimation fidelity, generalization to unseen networks, pre-verification efficiency, and the contribution of individual design components. Experiments run on a laptop (Intel Core i7-10510U) and a workstation (NVIDIA RTX A6000).
\vspace{-0.2cm}
\subsection{Experiment Setup}
\label{sec:ExperimentSetup}
\begin{figure}[!t]
    \centering    \includegraphics[width=1\linewidth]{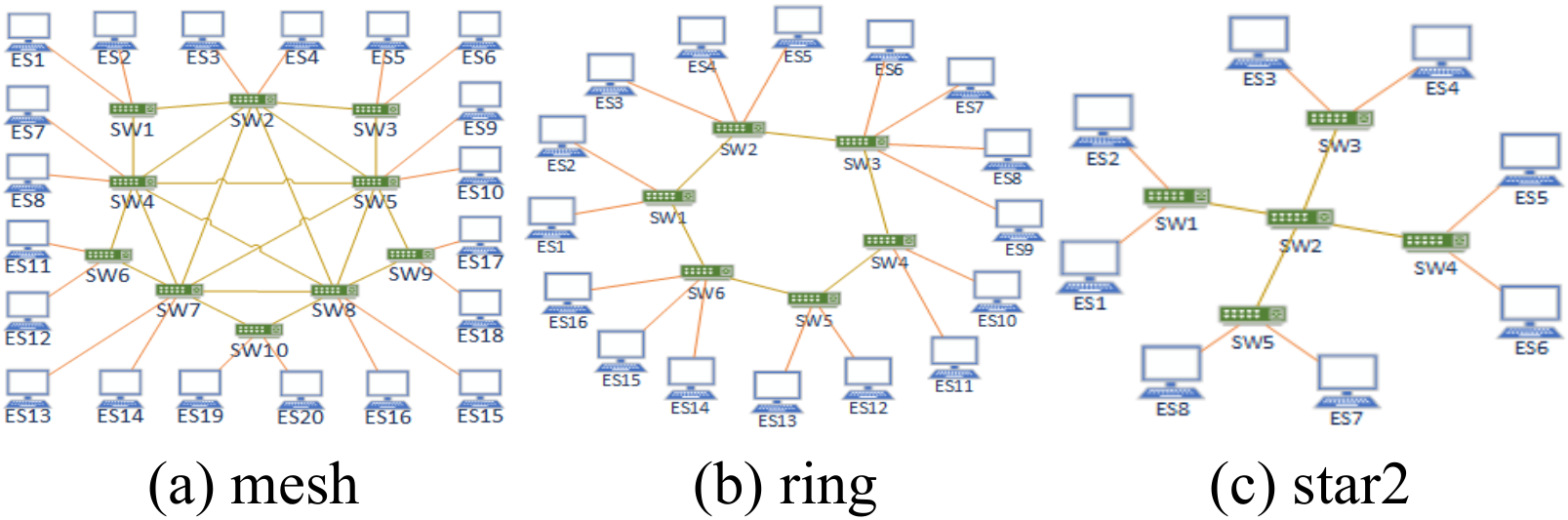}
    \vspace{-0.5cm}
    \caption{Synthetic Topologies.}
    \label{fig:topology}
    \vspace{-0.4cm}
\end{figure}

\begin{figure}[!t]
    \centering    \includegraphics[width=0.8\linewidth]{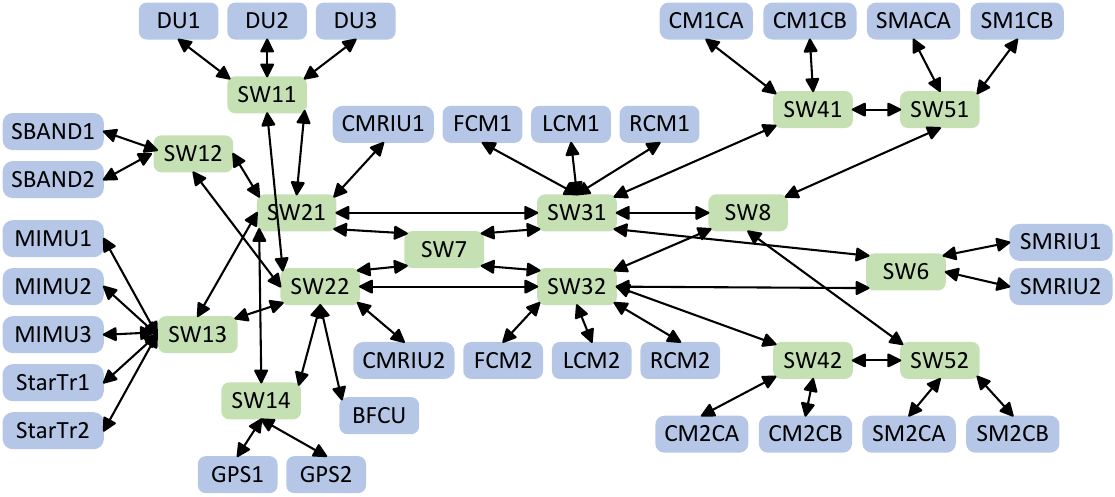}
    \vspace{-0.2cm}
    \caption{Realistic Test Case Orion CEV Topology.}
    \label{fig:Orion}
    \vspace{-0.2cm}
\end{figure}

\vspace{-0.1cm}
To evaluate scalability across different network structures, we design four synthetic topologies: \textbf{mesh}, \textbf{ring}, and \textbf{star2} (shown in Fig.~\ref{fig:topology}), along with an extended variant of \textbf{star2}, denoted as \textbf{star4}, obtained by adding two more ESs per branch of \textbf{star2}. Combined with the realistic industrial-scale test case, Orion CEV~\cite{orion} (Fig.~\ref{fig:Orion}), these five scenarios are used for both training and testing. For the synthetic topology dataset, we also generate network instances of varying sizes (ranging from 10 to 100 devices) for generalization experiments. Furthermore, to test generalization on unseen industrial networks, we additionally introduce three topologies from automotive Original Equipment Manufacturers: Renault (\textbf{rn}), Volvo (\textbf{vlv}), and FACE (\textbf{face})~\cite{navet2018automating,navet2019early,navet2020early}, which are used exclusively in the screening threshold analysis (Sec.~\ref{sec:acceleration-performance}).
For traffic distribution, we generate both TT and ET flows randomly, following the IEEE 802.3 standard. For each flow, the source and destination ESs are selected uniformly at random from all terminals. Frame lengths range from 64 to 1518 bytes. Periods for TT flow are randomly chosen from $\{5, 10, 15, 20, 30\} ms$ and from $\{5, 10\} ms$ for ET flows. ET flow priorities are randomly assigned from $\{1, 2, 3\}$. In the default setting, each topology contains 24 to 48 flows.

For baseline comparison, we benchmark DeepNC against two SOTA ML baselines, both adapted specifically for our delay prediction task. For the \textbf{Mai} baseline~\cite{mai2022gnn}, originally for binary schedulability classification, we adapt its readout layer to output continuous delay bounds. 
For the \textbf{RouteNet}~\cite{ferriol2023routenetfermi} baseline, based on the well-known RouteNet-Fermi architecture initially designed to predict simulation-based network performance (e.g., via OMNeT++), we repurpose it by retraining the model using the WCDs calculated by NC as ground-truth labels. To accommodate the hybrid TSN/TAS+CBS network, we also feed the TAS-related and CBS-related parameters extracted in Sec.~\ref{sec:FeatueExtraction} as input features to both baselines. To ensure a fair comparison, all baselines adopt the same training configuration as DeepNC.
We implement DeepNC with TensorFlow, using T=8 message-passing iterations and a hidden dimension of 32. All MLPs have 2 layers with the same hidden size. We use ReLU~\cite{2011relu} activation, MAE loss, and Adam~\cite{2014adam} optimizer with learning rate 0.001. All models are trained for 100 epochs (300 steps each).

\vspace{-0.2cm}

\subsection{Estimation Fidelity Comparison}
\label{sec:overall-performance}
\vspace{-0.1cm}
We compare DeepNC against two baselines across four topologies. For each topology, we use 600 training, 200 validation, and 200 test samples. Performance is measured using Mean Absolute Percentage Error (MAPE) and the coefficient of determination ($R^2$).

As shown in Table~\ref{tab:mape_r2}, DeepNC significantly outperforms both baselines on all test sets. Specifically, DeepNC achieves an average MAPE of 8.36\%, reducing errors by 61.96\% and 68.68\% compared to Mai and RouteNet, respectively. Furthermore, DeepNC maintains a consistently high average $R^2$ of 0.9547, whereas the baselines exhibit significant instability.

\begin{table}[!t]
\vspace{-0.1cm}
\caption{Comparison of Regression Accuracy}
\vspace{-0.2cm}
\centering
\setlength{\tabcolsep}{6pt}
\small 
\resizebox{1\linewidth}{!}{
\begin{tabular}{l cc cc cc}
\toprule
\textbf{Dataset} & \multicolumn{2}{c}{\textbf{Mai}} & \multicolumn{2}{c}{\textbf{RouteNet}} & \multicolumn{2}{c}{\textbf{DeepNC}} \\
\cmidrule(lr){2-3} \cmidrule(lr){4-5} \cmidrule(lr){6-7}
& \textbf{MAPE} & $R^2$ & \textbf{MAPE} & $R^2$ & \textbf{MAPE} & $R^2$ \\
\midrule
\textbf{mesh} & 38.53\% & 0.5656 & 26.52\% & 0.6060 & \textbf{5.62\%} & \textbf{0.9768} \\
\textbf{ring} & 25.88\% & 0.7214 & 37.12\% & 0.4183 &  \textbf{6.90\%} & \textbf{0.9692} \\
\textbf{star2} & 18.20\% & 0.7021 & 21.99\% & 0.5969 & \textbf{6.47\%} & \textbf{0.9524} \\
\textbf{orion} & 19.16\% & 0.7228 & 25.75\% & 0.7421 & \textbf{14.44\%} & \textbf{0.9202}  \\
\bottomrule
\end{tabular}
}
\label{tab:mape_r2}
\vspace{-0.1cm}
\end{table}

To validate the applicability of DeepNC in industrial environments, we also analyzed its performance on the Orion CEV network. This real-world test case features highly irregular connectivity and heavier traffic flow. As shown in Table~\ref{tab:mape_r2}, despite the challenges posed by limited training data for this specific topology, DeepNC still maintained a high R² value of 0.9202. This result is significant for industrial deployments: it demonstrates that because DeepNC learns the underlying analytical logic of NC rather than memorizing the topology, it maintains high prediction fidelity even in complex real-world scenarios with scarce training data.

\begin{figure}[!t]
    \centering    \includegraphics[width=1\linewidth]{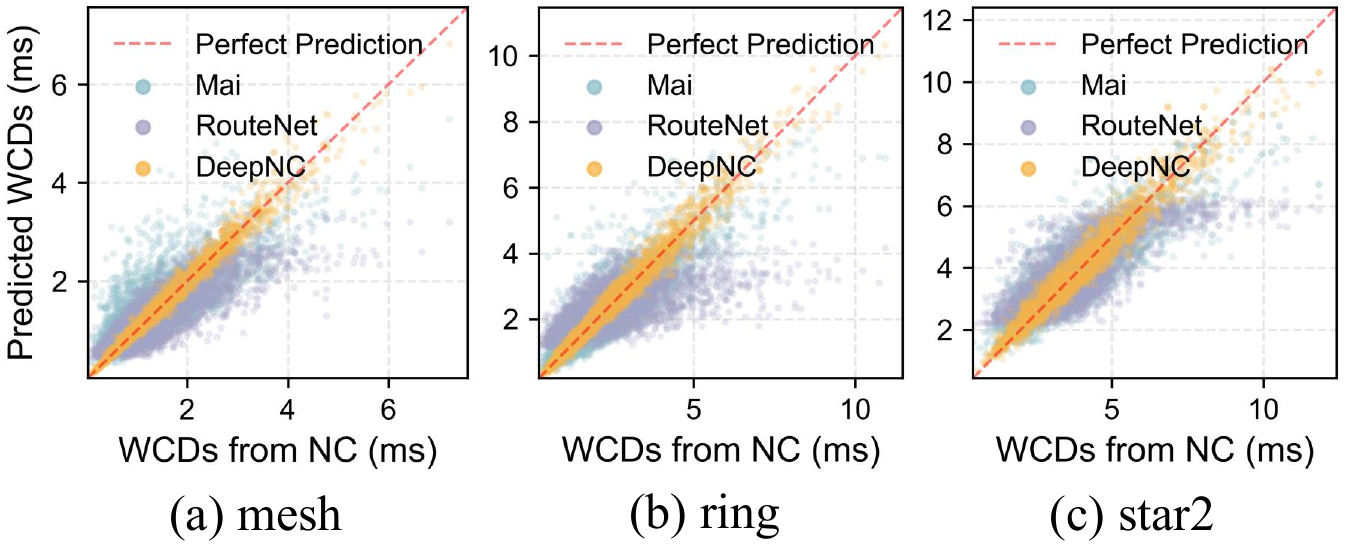}
    \vspace{-0.6cm}
    \caption{Learning-based vs. formal NC-computed WCDs.}
    \label{fig:fourtest}
    \vspace{-0.2cm}
\end{figure}

Fig.~\ref{fig:fourtest} shows scatter plots for three topologies, comparing WCDs predictions of each learning-based model with the NC-computed ground truth\footnote{DeepNC serves as a surrogate within the formal NC verification workflow rather than replacing it. It is meaningful as a surrogate only if it approximates the analytical results of NC, regardless of the pessimism of those results.}. The predictions of DeepNC lie tightly along the diagonal ($x=y$), indicating high fidelity to the NC bounds. In contrast, Mai and RouteNet show larger deviations.
\vspace{-0.1cm}

\vspace{-0.5cm}
\subsection{Generalization Performance Evaluation}
\label{sec:Scalability&RobustnessAnalysis}
\begin{figure}[!t]
    \centering
    \includegraphics[width=1\linewidth]{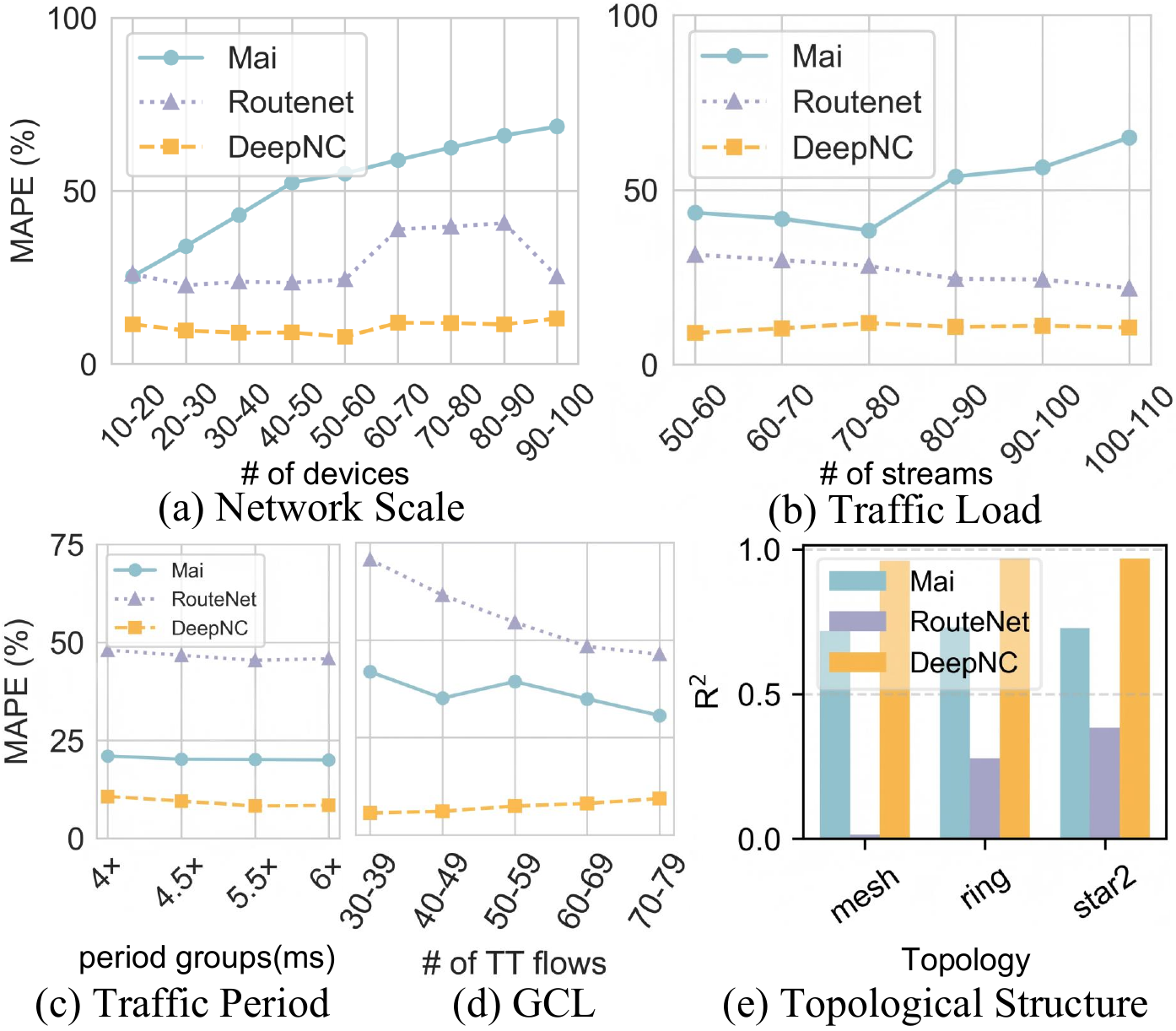}
    \vspace{-0.6cm}
    \caption{Comparison of Scalability and Robustness.}
    \label{fig:generalization}
    \vspace{-0.2cm}
\end{figure}

\vspace{-0.1cm}
We evaluate the adaptability and stability of DeepNC on unseen networks across five dimensions: network scale, traffic load, traffic parameters, GCL configurations, and topological structure. The results are summarized in Fig.~\ref{fig:generalization}.

For network scale, we train all models (DeepNC, Mai, and RouteNet) exclusively on the 13-device \textbf{star2} topology as shown in Fig.~\ref{fig:topology}(c), and test them on a set spanning a wide range of network sizes (10 to 100 devices). As shown in Fig.~\ref{fig:generalization}(a), DeepNC maintains a stable MAPE below 14\% across all scales. While its error shows a slight upward trend, the overall predictive performance remains robust. In contrast, the error for Mai grows linearly with network size, and the error for RouteNet is highly unstable.

For traffic load, we train all models on the \textbf{star2} topology with 24–48 flows and evaluate them on configurations with a much higher flow count (50 to 110 flows). As shown in Fig.~\ref{fig:generalization}(b), DeepNC consistently maintains an MAPE below 12\% across all tested flow densities. In comparison, both baselines lag significantly behind DeepNC. This again highlights the limitation of the baseline models: they lack a structural awareness of the NC process, reflecting overfit to the specific network characteristics of the training data.

For traffic parameters, we train all models on \textbf{star2} with TT periods from $\{5, 10, 15, 20, 30\}ms$ and ET periods from $\{5, 10\} ms$. Test sets use periods as multiples of 4, 4.5, 5.5, and 6 $ms$. As shown in Fig.~\ref{fig:generalization}(c), DeepNC achieves MAPE between 8.29\% and 10.69\% (avg. 9.2\%), while Mai ($\approx$ 20.3\%) and RouteNet ($\approx$ 46.5\%) lag far behind. This indicates that DeepNC learns the underlying NC logic rather than memorizing specific period values.

For GCL configurations, we train all models on \textbf{star2} with 15–30 TT flows and test them on five groups: 30–39, 40–49, 50–59, 60–69, and 70–79 TT flows. As shown in Fig.~\ref{fig:generalization}(d), DeepNC achieves MAPE between 5.67\% and 9.37\% (avg. 7.35\%), with only a slight degradation as TT flows increase, while Mai ($\approx$ 36.4\%) and RouteNet ($\approx$ 56.3\%) perform significantly worse. This confirms that DeepNC provides accurate WCDs predictions for ET flows under unseen GCL configurations in TSN/TAS+CBS networks.

For topological structures, we train all models on a single topology (e.g., \textbf{mesh}) as described in Sec.~\ref{sec:overall-performance}, and then test them on a mixed set comprising the remaining two topologies (e.g., \textbf{ring} and \textbf{star2} when training on \textbf{mesh}), with balanced samples from each topology. As shown in Fig.~\ref{fig:generalization}(e), DeepNC demonstrates stable cross-topology performance, consistently achieving $R^2$ scores above 0.96 regardless of the training topology. In contrast, the Mai baseline shows only moderate adaptability, while RouteNet performs poorly. This is because DeepNC internalizes the structural dependencies of NC rather than overfitting to specific topologies.

\vspace{-0.2cm}
\subsection{Pre-verification Performance of DeepNC}
\label{sec:acceleration-performance}
\begin{figure}[!t]
    \centering
    \includegraphics[width=1\linewidth]{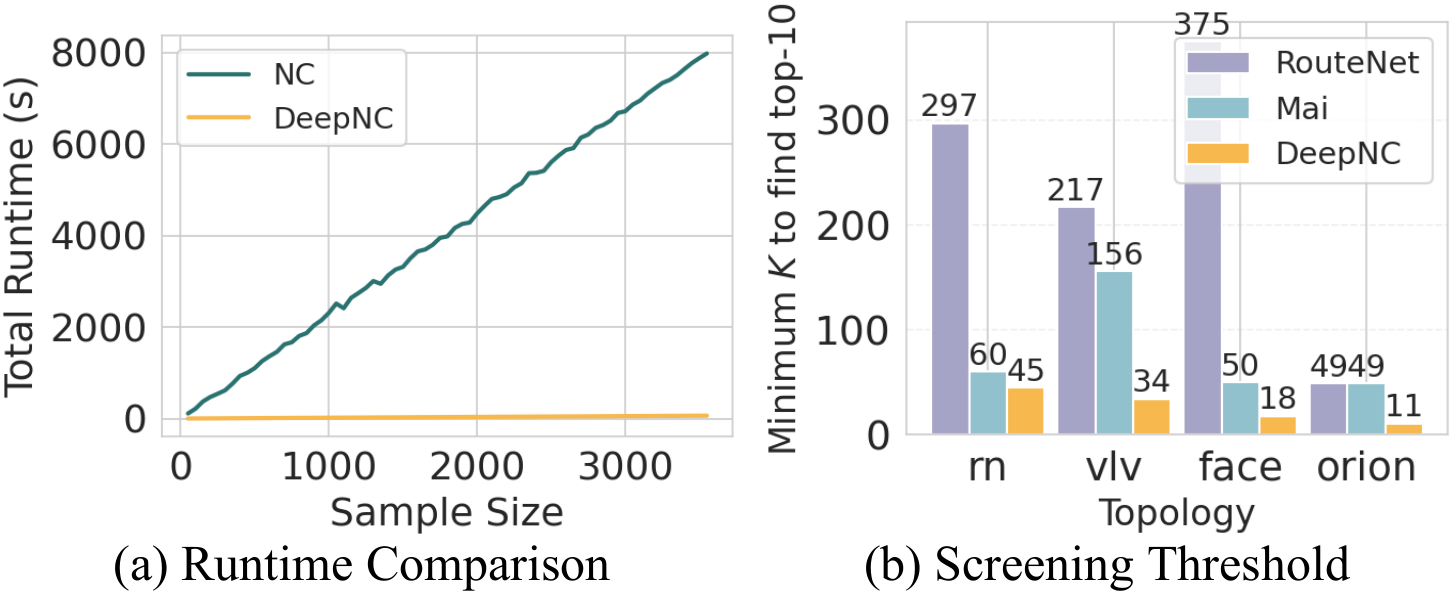}
    \vspace{-0.6cm}
    \caption{Pre-verification performance.}
    \label{fig:accel}
    \vspace{-0.4cm}
\end{figure}

\vspace{-0.1cm}
This subsection evaluates DeepNC as a verification accelerator for configuration-space exploration in terms of runtime and screening capability. We compare against the TFA-based NC tool~\cite{ZhaoTIE} (min-plus operators in RTC toolbox~\cite{RTCtoolbox}) on 3,582 configurations with varying candidate set sizes. As shown in Fig.~\ref{fig:accel}(a), DeepNC is two orders of magnitude faster than NC formal analysis, with near-constant runtime via parallel batch inference versus linear growth for NC. The gap widens as candidate sets grow, confirming DeepNC as an ultra-fast pre-verification surrogate that reserves formal verification for a small promising subset. We further evaluate fidelity of DeepNC in configuration ranking relative to formal NC on four unseen real-world topologies (\textbf{rn}, \textbf{vlv}, \textbf{face} and \textbf{orion}), with 400 random configurations per topology (Sec.~\ref{sec:ExperimentSetup}). The maximum predicted rank among the top-10 true optimal configurations, as determined by exact formal NC, defines the minimum screening threshold $K$ to guarantee no miss. Smaller $K$ means fewer candidates for expensive formal analysis. As shown in Fig.~\ref{fig:accel}(b), DeepNC achieves the smallest $K$ across all topologies, reducing the number of formal verification calls by 93.25\% on average, while also outperforming baselines with a reduction of 73.3\% relative to them. Overall, DeepNC delivers high-speed inference and reliable ranking, reducing verification overhead in configuration-space exploration.
\begin{figure}
    \centering    \includegraphics[width=0.9\linewidth]{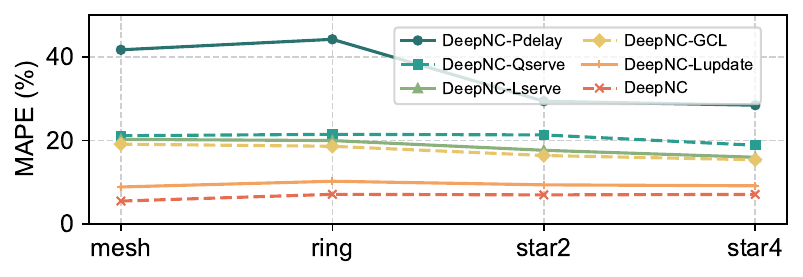}
    \vspace{-0.3cm}
    \caption{Ablation Results for Different DeepNC Variants.}
    \vspace{-0.2cm}
    \label{fig:ablation}
\end{figure}
\vspace{-0.1cm}
\subsection{Ablation Experiments}
\vspace{-0.1cm}
To analyze the contribution of each design component in DeepNC, we conducted an ablation study. We compared the full model against five variants, each with a particular module removed or modified. \textbf{DeepNC-GCL} removes GCL features entirely. \textbf{DeepNC-Pdelay} alters the delay readout mechanism by directly predicting end-to-end worst-case delay from \myconcept{Flow Node} state. 
\textbf{DeepNC-Lupdate} disables the iterative update for \myconcept{Link Node} states. 
\textbf{DeepNC-Lserve} excludes the \myconcept{Link Node} state from the service hidden state synthesis. 
\textbf{DeepNC-Qserve}, in contrast, excludes the \myconcept{Queue Node} state from the service hidden state synthesis.
As shown in Fig.~\ref{fig:ablation}, all ablated variants perform worse than the complete DeepNC model, confirming that each component is critical for overall performance. \textbf{DeepNC-Pdelay} suffers the most severe degradation (MAPE $>$ 40\%), as it discards intermediate hop-by-hop NC emulation information, leading to prediction collapse.

\section{Conclusion}
\label{sec:Conclusion}
\vspace{-0.1cm}
We presented DeepNC, a novel pre-verification surrogate module for TSN configuration-space exploration. Rather than replacing formal verification, DeepNC acts as a high-speed pre-verification filter that rapidly eliminates unpromising configurations while reserving rigorous NC analysis for a small set of promising candidates. By structurally integrating the analytical principles of TFA-based Network Calculus into GNN, DeepNC enables high-fidelity worst-case delay regression for hybrid TSN/TAS+CBS architectures. Although this work is currently limited to TFA-based NC analysis for hybrid TSN/TAS+CBS networks, the proposed structural integration paradigm can be easily extended to other TFA-based TSN architectures through corresponding graph abstractions and model retraining. Additionally, extending this paradigm to other NC analyses, such as SFA and PMOO, constitutes an important direction for future work.

\bibliographystyle{IEEEtran}  
\bibliography{refs1}        

\end{document}